\documentclass[aps, pra, superscriptaddress, twocolumn]{revtex4-1}

\usepackage{amsmath}
\usepackage{amssymb}
\usepackage{graphicx}
\usepackage{subfigure}
\usepackage{color}

\newcommand{\dd}{\text{d}}

\newcommand{\modulus}[1]{\left| #1 \right|}
\newcommand{\intinfty}{\int_{- \infty}^{\infty}}

\newcommand{\commentout}[1]{}

\begin{document}

\title{Ground-state electronic structure of quasi-one-dimensional wires in semiconductor heterostructures}
\author{E. T. Owen}
\email{e.owen@hw.ac.uk}
\affiliation{London Centre for Nanotechnology, 17-19 Gordon Street, London WC1H 0AH, United Kingdom}
\affiliation{Cavendish Laboratory, Department of Physics, University of Cambridge, Cambridge CB3 0HE, United Kingdom}
\affiliation{Institute of Photonics and Quantum Sciences, Heriot-Watt University Edinburgh EH14 4AS, United Kingdom}
\author{C. H. W. Barnes}
\affiliation{Cavendish Laboratory, Department of Physics, University of Cambridge, Cambridge CB3 0HE, United Kingdom}

\date{\today}

\begin{abstract}

We apply density functional theory, in the local density approximation, to a quasi-one-dimensional electron gas in order to quantify the effect of Coulomb and correlation effects in modulating, and therefore patterning, the charge density distribution.  Our calculations are presented specifically for surface-gate-defined quasi-one-dimensional quantum wires in a GaAs-AlGaAs heterostructure but we expect our results to apply more generally for other low dimensional semiconductor systems.  We show that at high densities with strong confinement, screening of electrons in the direction transverse to the wire is efficient and density modulations are not visible.  In the low-density, weak-confinement regime, the exchange-correlation potential induces small density modulations as the electrons are depleted from the wire.  At the weakest confinements and lowest densities, the electron density splits into two rows thereby forming a pair of quantum wires that lie beneath the surface gates.  An additional double-well external potential forms at very low density which enhances this row splitting phenomenon. We produce phase diagrams that show a transition between the presence of a single quantum wire in a split-gate structure and two quantum wires.  We suggest that this phenomenon can be used to pattern and modulate the electron density in low-dimensional structures with particular application to systems where a proximity effect from a surface gate would be valuable.  

\end{abstract}


\maketitle

\section{Introduction}

Very large scale integration and the invention of the microprocessor has revolutionized almost every aspect of the modern world.  The importance of the miniaturization of the field effect transistor (FET) is most dramatically demonstrated by the celebrated Moore's Law, whereby the size and cost of transistors decreases at an exponential rate.  However, since its inception, it has been known that Moore's Law will inevitably fail at the atomic level as quantum effects become dominant and the conventional operation mode of FETs breaks down.  Understanding and manipulating quantum effects in these devices will play an essential role in extending Moore's law into the twenty-first century.

In semiconductor heterostructures, the FET is realised through a combination of band-gap engineering and electrostatic gating.  Dopants are used to promote electrons into the material's conduction band whilst patterned metallic surface gates define electrostatic potentials which alter the density and the dimensionality of the electron system~\cite{Thornton:1986, Laux:1988}.  Using these techniques, the electrons can be confined to structures with features which are on the same length scale as the surface gate features.

In conventional devices, engineered electrostatic potentials are the only method for confining the electrons.  However, at low densities electron screening becomes less efficient and Coulombic repulsion can be used in concert with electrostatic confinement to generate features at smaller length scales.  
For example, whilst single electrons can be localised into a set of coupled quantum dots using sophisticated gate arrays~\cite{Shulman:2012, Thalineau:2012, Puddy:2015}, spontaneous localisation of the electron gas into a Wigner crystal state provides a simpler alternative for generating an ordered set of electrons.  
Quantum Monte Carlo~\cite{Guclu:2009} and density functional theory~\cite{Berggren:2002, Jaksch:2006, Welander:2010} simulations have shown that, within a quasi-one-dimensional system, interactions cause the electrons to localise and form a one-dimensional Wigner crystal.  As the confining potential is weakened, the crystal undergoes a phase transition first to zig-zag, then double row configurations~\cite{Welander:2010, Mehta:2013, Silvi:2014}.  
Experimental evidence of the transition to a localised electron state has been observed in electron transport measurements in GaAs/AlGaAs heterostructures.  A quasi-one-dimensional wire can be defined using metallic surface gates~\cite{Wharam:1988, vanWees:1988} and an additional top gate can be used to vary the electron density in the wire~\cite{Hew:2009, Smith:2009, Kumar:2014}.  In the high density regime, the electrons behave as non-interacting particles and the conductance is quantised in units of $2 e^2 / h$~\cite{Buettiker:1990} but as the electron's density in the wire is reduced, the first plateau suddenly disappears.  Further reduction in the density leads to a revival of this plateau.  It has been speculated that this ``missing plateau'' can been attributed to the degeneracy of the zig-zag state~\cite{Smith:2009}.
The electron separation in these devices is considerably smaller than could be achieved using surface gate patterning.

In this paper, we propose a method for exploiting Coulombic repulsion in order to create two quantum wires in the electron gas with a separation width similar to the length scale of the surface gate pattern.  We use density functional theory (DFT) in the local density approximation (LDA) to find the ground-state electronic structure of translationally invariant quantum wires with experimentally realistic, transverse confining potentials.  The confining potentials are calculated exactly, including all anharmonic and image charge contributions, by solving Poisson's equation using finite-element analysis (FEA) with adaptive gridding~\cite{Bangerth:2015}.  These computational techniques allow us to probe a regime where electrons are confined through a combination of electrostatic and Coulombic potentials, which opens the door to a wide variety of applications in future quantum one-dimensional technologies.  

We start by showing that in the strong-confinement regime, electron screening almost completely removes transverse density modulations.  These results contradict the observations of Laux \emph{et al.}~\cite{Laux:1988} and are due to the highly accurate FEA potentials used.  We extend these results into the low-density, weak-confinement regime and find that transverse density modulations exist at low densities for a range of confinements.  These modulations are only observed in the LDA and therefore are driven by exchange-correlation effects.  For the weakest possible confinement and the lowest achievable densities, the electrostatic repulsion of the electrons splits the density into two rows.  For the experimental setup used to reach this regime, it is possible for an external double-well potential to form as the top gate is negatively biased with respect to the side gates.  Our simulations show that this does occur but only after a double-row configuration of the density has appeared due to Coulombic interactions.

As the rows decouple, the energy difference between the ground and first excited sub-bands of the wire vanishes and the first conductance plateau disappears.  It is possible that disorder could break this degeneracy as the confinement potential weakens, which would lead to the revival of this plateau as measured in Refs.~\cite{Hew:2009, Smith:2009, Kumar:2014}, but this effect is not included.  The results demonstrate that anharmonic contributions to the confinement potential are crucial to understanding the electronic structure of weakly-confined one-dimensional wires.  Electron localisation in this regime is strongly determined by the interplay between electron-electron and external confinement potentials.

The paper is organized as follows:  In Sec.~\ref{sec:model}, we describe the model used to calculate the ground states density of quasi-one-dimensional quantum wires with arbitrary semiconductor heterostructures and surface gate geometries.  We use this model in Sec.~\ref{sec:LS} to show that in the strong-confinement regime electron screening is efficient and density modulations effectively disappear.  In Sec.~\ref{sec:phase_diagram}, we explore the electronic structure in the low-density, weak-confinement regime and we report the various phases of the quasi-one-dimensional wire.  We present our conclusions in Sec.~\ref{sec:conclusions} along with possible experimental methods for measuring the density modulations predicted in this paper.

\section{Model}
\label{sec:model}

Semiconductor heterostructures consist of layers of semiconductor material with differing alloy composition, typically grown by molecular beam epitaxy, which produce a variable band-gap material.  Ionized dopants introduced into specific layers generate internal electric fields which can bend the conduction band at a remote interface below the chemical potential and allow low-dimensional electron systems to form.  Additional fields can be applied using metallic surface gates to shape the electrons into quasi-one-dimensional quantum wires.  In order to accurately calculate the structure of the electron density for these low-dimensional systems, we will include the effects of dopant densities, surface charges and biased metallic surface gates on arbitrary layered semiconductor heterostructures.  By simulating the potential from the entire device, and not resorting to analytic models, anharmonic contributions present in the confining potential will be included which can strongly affect the form of the electron density.

Firstly, we must calculate the electrostatic potential $\phi (\vec{r})$.  For a given charge density $\rho (\vec{r})$, $\phi (\vec{r})$ is the solution to Poisson's equation
\begin{equation}
  \label{eq:Poisson}
  g \left[\phi (\vec{r}), \rho (\vec{r})\right] \equiv - \nabla \cdot \left(\epsilon (\vec{r}) \nabla \phi (\vec{r}) \right) - \frac{\rho (\vec{r})}{\epsilon_0} = 0
\end{equation}
where $\epsilon (\vec{r})$ is the relative permittivity~\footnote{The band gap and relative permittivities of GaAs and AlGaAs are taken from the IOFFE semiconductor database http://www.ioffe.ru/SVA/NSM/Semicond/index.html.  The relative permittivity of the insulating polymer (PMMA) was set to $\epsilon_{r, \mathrm{PMMA}} = 2.6$.} of the layered semiconductor materials at $\vec{r} = (x, y, z)$.  The boundary conditions on $\phi (\vec{r})$ are determined by the voltages on the metallic surface gates.   For free surfaces, we apply von Neumann boundary conditions $\vec{n} \cdot \nabla \phi (\vec{r}) = 0$ where $\vec{n}$ is the normal to the domain boundary.  We calculate the potential using finite-element analysis (FEA) on an adaptively refined mesh using the deal.II finite-element library~\cite{Bangerth:2015} which increases the computational efficiency and accuracy of the solution as well as eliminating high-order Fourier components in $\phi (\vec{r})$ which may be generated by the discretisation onto a regular grid.

Secondly, we need to calculate the charge density $\rho (\vec{r})$ for an arbitrary semiconductor heterostructure with an electrostatic potential $\phi(\vec{r})$.  We start by separating $\rho (\vec{r})$ into static and mobile charges.  Our method for calculating the static charges -- which consist of the dopant and surface charges -- is described in App.~\ref{app:dopant_calc}.  The mobile charges considered in this paper are electrons confined to a narrow region below a GaAs/AlGaAs interface.  To calculate this density, we need to solve the quantum many-body problem, which is impossible, so approximations need to be made.  The approach used in this paper is to use density functional theory (DFT) within the Hartree and local density approximations (LDA).

The electron charge density $\rho_e (\vec{r}) = - e n (\vec{r}) = - e \sum_j \modulus{\Psi_j (\vec{r})}^2$ is calculated by solving the Kohn-Sham~(KS) equations in the effective mass approximation
\begin{eqnarray}
  \left( -\frac{\hbar^2 \nabla^2}{2 m_*} + V_{ext}(\vec{r}) + V_{H}[n](\vec{r}) + V_{xc}[n] (\vec{r}) \right) \Psi_j (\vec{r}) \nonumber \\
  \label{eq:KS_Hamiltonian}
  = E_j \Psi_j (\vec{r})
\end{eqnarray}
where $m_* = 0.067 m_e$ is the reduced mass of the electron in GaAs.  The external potential $V_{ext}(\vec{r})$ generated by the dopants, surface charges and applied metallic gate bias voltages and the classical electron-electron or Hartree potential $V_{H}[n](\vec{r})$ are both obtained from the solution to Eq.~\ref{eq:Poisson}
\begin{equation}
  V_{\mathrm{ext}}(\vec{r}) + V_{H}[n](\vec{r}) = \frac{1}{2} E_g (\vec{r}) - e \phi(\vec{r})
\end{equation}
where, for layered semiconductor heterostructures, the band gap $E_g (\vec{r})$ is a function of the growth direction $z$ only. $V_{xc}[n](\vec{r})$ is the exchange-correlation potential which contains all quantum many-body effects.  For the Hartree approximation $V_{xc} [n] (\vec{r}) = 0$ whilst in the LDA we use the parameterization of Perdew and Wang~\cite{Perdew:1992} in effective atomic units with 1 Ry$^* = m_* e^4 / 2 \hbar^2 \epsilon_0^2 \epsilon_r^2$ and 1 $a_B^* = \hbar^2 \epsilon_0 \epsilon_r / m_* e^2$.

One-dimensional devices are translationally invariant in the longitudinal direction $x$ so the solutions to the Kohn-Sham equation are plane waves
\begin{equation}
  \Psi_j(\vec{r}) = \psi_j(y, z) e^{i k x}
\end{equation}
Inputting $\Psi_j(\vec{r})$ into Eq.~\ref{eq:KS_Hamiltonian} gives
\begin{eqnarray}
  \left(- \frac{\hbar^2}{2 m_*} \left( \frac{\partial^2}{\partial y^2} + \frac{\partial^2}{\partial z^2} - k^2 \right) + V(y, z) \right) \psi_j (y, z) \nonumber \\
  \label{eq:2D_Hamiltonian}
  = E_j \psi_j (y, z)
\end{eqnarray}
where $V(y, z) = E_g(z) / 2 - e \phi(y, z) + V_{xc}[n](y, z)$.  The transverse wave functions $\psi_j (y, z)$ are independent of $k$ so the parabolic dispersion relation for each of the one-dimensional sub-bands $j$ can be integrated numerically to give the electron density:
\begin{equation}
    \label{eq:2D_Density}
    n(y, z) = \sum_j \int_{-\infty}^{\infty} f(E, T) g_{1D} (E; \varepsilon_j) \modulus{\psi_j(y, z)}^2 \dd E
\end{equation}
where $\varepsilon_j = E_j - \hbar^2 k^2 / 2 m_*$, $g_{1D} (E; \varepsilon_j) = \frac{1}{\pi \hbar} \sqrt{2 m_* / (E - \varepsilon_j)}$ is the one-dimensional density of states and $f(E, T) = (1 + e^{E / k_B T})^{-1}$ is the Fermi-Dirac function for the system at temperature $T$.

Finally, we use an iterative procedure to find a consistent solution for Eqs.~\ref{eq:Poisson} and~\ref{eq:KS_Hamiltonian}.  An initial electrostatic potential $\phi_0$ is calculated for the system with $n (\vec{r}) = 0$.  This potential is updated using an adaptive Newton step method $\phi_{n+1} = \phi_{n} + t_n x_n$ where $t_n$ is a mixing parameter and the Newton step $x_n$ is the solution to $g'(\phi_n) x_n = - g(\phi_n)$ where $g(\phi_n) \equiv g\left[\phi_n (\vec{r}), \rho(\phi_n (\vec{r}'))(\vec{r})\right]$ and $g'(\phi_n)$ is the Jacobian of $g(\phi_n)$.  Following Laux and Stern~\cite{Laux:1986}, we assume that
\begin{align}
  \frac{\partial \rho_e(\phi_n(\vec{r}))}{\partial \phi_n(\vec{r}')} \approx & e \sum_j \modulus{\psi_j (y, z)}^2 \nonumber \\
   & \quad \cdot \frac{\partial}{\partial \phi} \int_{-\infty}^{\infty} f(E, T) g_{1D} (E; \varepsilon_j) \dd E
\end{align}
The Newton step is solved using the same FEA methods as Eq.~\ref{eq:Poisson}.  The mixing parameter $t_n$ is calculated by finding the zero of the function $F(t) = g^T(\phi_n + t x_n) \cdot x_n$ which accelerates convergence~\cite{Bank:1980}.  We update the exchange-correlation potential using a simple function mixing method
\begin{equation}
  V_{xc, \mathrm{new}}[n](\vec{r}) = (1 - \alpha) V_{xc, \mathrm{old}}[n](\vec{r}) + \alpha V_{xc, \mathrm{calc}}[n](\vec{r})
\end{equation}
where $\alpha = 0.3$ is a constant mixing parameter.  The iteration scheme terminates when the maximum change in the potential $V(y, z)$ for the grid on which the electron density is calculated is less than 0.1$\mu$eV.

To ensure that the converged values of $\phi(\vec{r})$ and $n (\vec{r})$ are consistent, the density obtained using this iterative procedure is used to calculate a new initial potential $\phi_0$.  Only when the maximum value of the initial Newton step $x_0$ is less than 4$\mu$eV do we consider the density to have reached convergence.

\section{Electron Densities in the Strong Confinement Regime}
\label{sec:LS}

\begin{figure*}
  \subfigure[][\label{fig:LS_Hartree_pot} Hartree]{\includegraphics[width = 8.5cm]{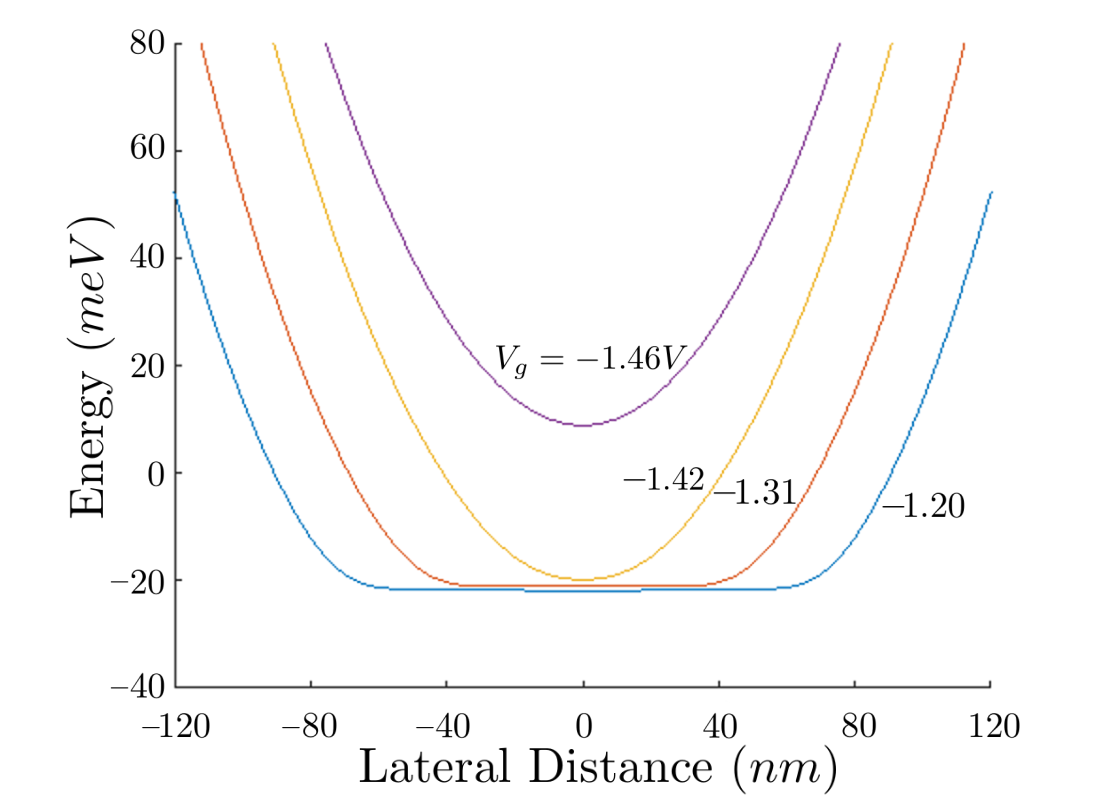}}
  \subfigure[][\label{fig:LS_LDA_pot} LDA]{\includegraphics[width = 8.5cm]{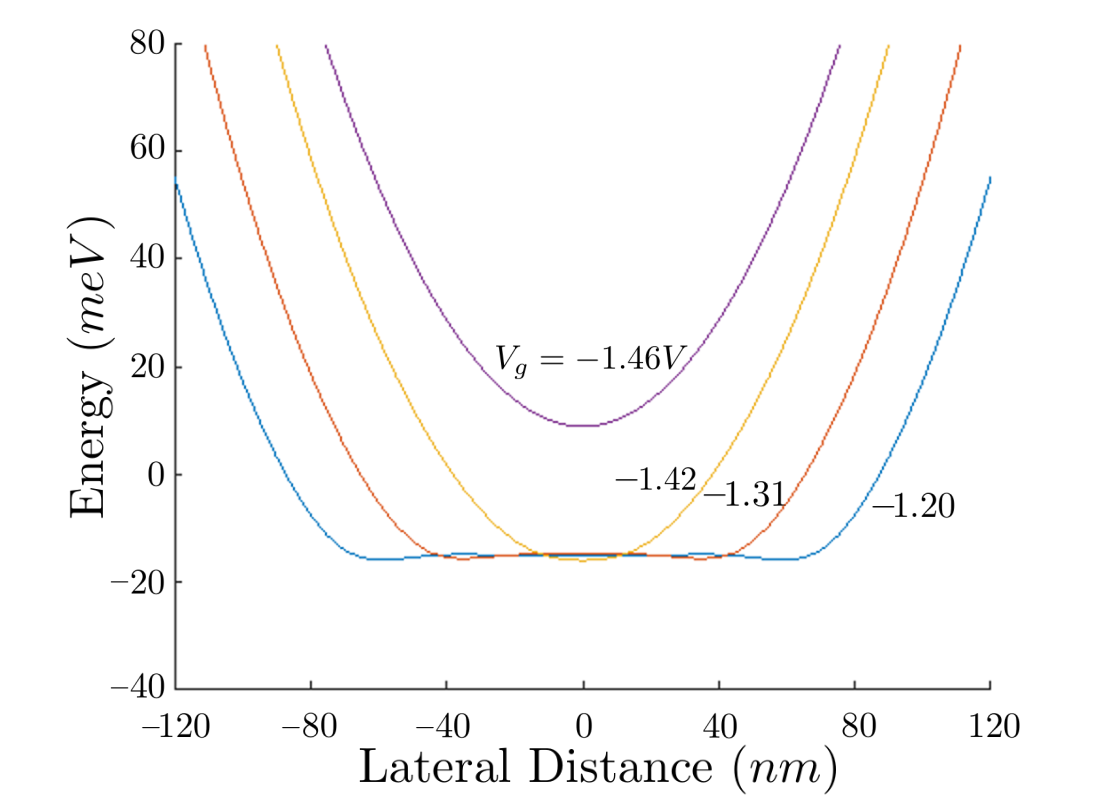}}
  \caption{(Color online) Lateral potential at 4.2K on a cross-section 6nm below the GaAs/AlGaAs interface for the device described in Sec.~\ref{sec:LS}.}
  \label{fig:LS_pot}
\end{figure*}

\begin{figure*}
  \subfigure[][\label{fig:LS_Hartree_dens} Hartree]{\includegraphics[width = 8.5cm]{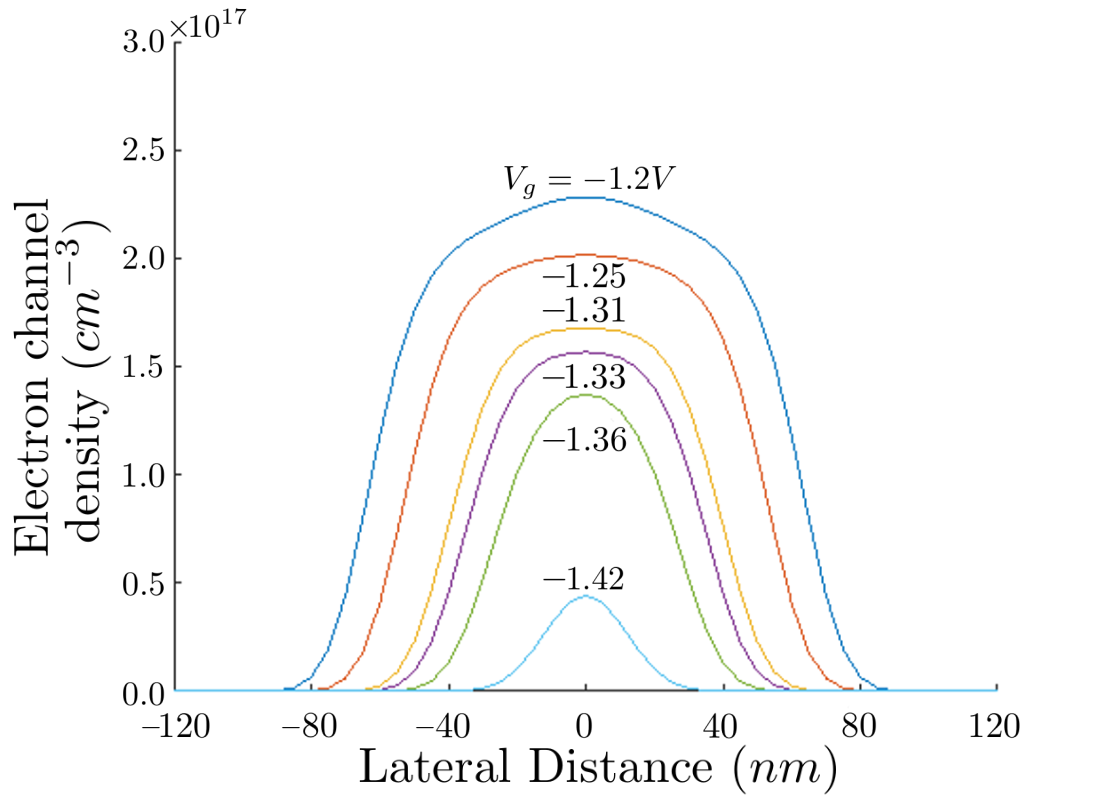}}
  \subfigure[][\label{fig:LS_LDA_dens} LDA]{\includegraphics[width = 8.5cm]{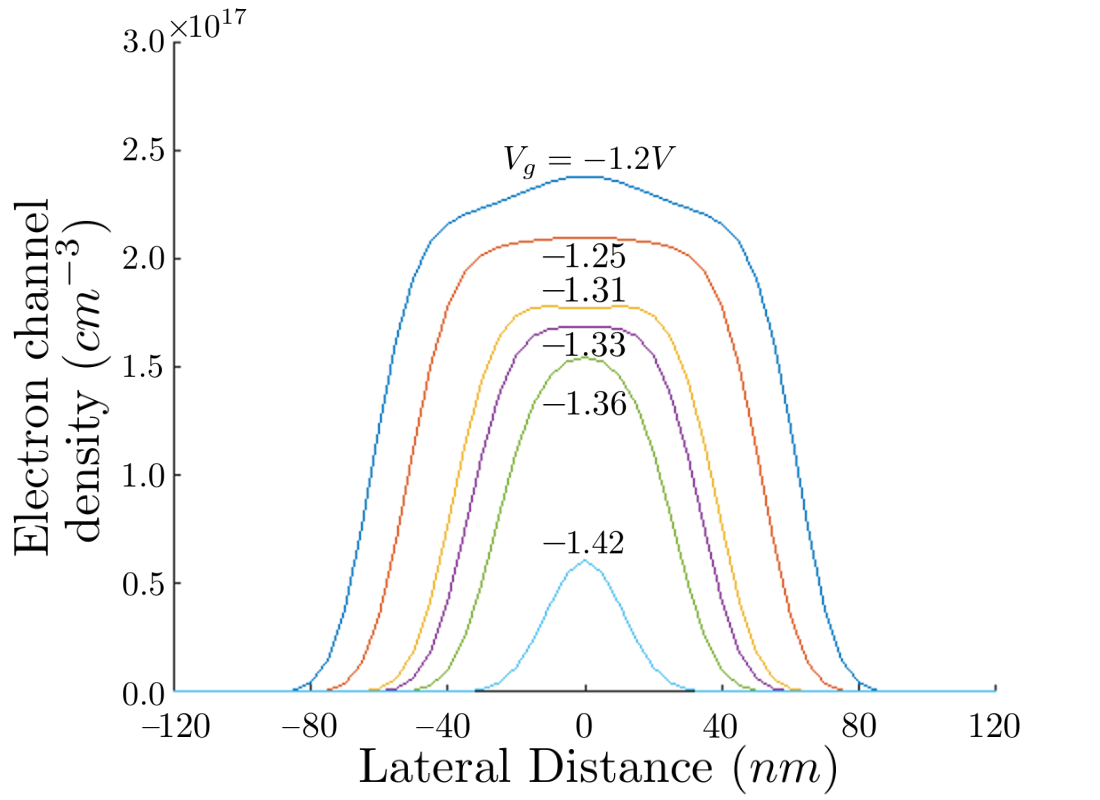}}
  \caption{(Color online) Electron charge density at 4.2K along the same cross-section as in Fig.~\ref{fig:LS_pot}.  The electron screening is efficient and density modulations in the transverse direction vanish for this strong confinement regime.  The difference between the Hartree and local-density approximations is minimal.}
  \label{fig:LS_dens}
\end{figure*}

For a truly one dimensional system, electrons are confined so that only motion in the longitudinal direction is possible.  However, in any realistic physical system, the confining potential is finite and the electrons will be able to move in the transverse direction which defines a quasi-one-dimensional wire.  We can define a strong confinement regime, where the electrons are forced into the center of the wire and there is no irregular structure of the electrons.  The electrons will spread out to some extent in the transverse direction as the density increases but, when the confinement is strong, screening is expected to be efficient so any density modulations should be washed out.

In semiconductor heterostructure devices, a pair of parallel negatively-biased metallic gates above an intrinsic two-dimensional electron gas will deplete the electrons underneath and confine them into a quasi-one-dimensional quantum wire.  Such a device was modelled by Laux \emph{et al.}~\cite{Laux:1988} and is described in detail in App.~\ref{subapp:LS_device}.  In this section, we reproduce their results and show that screening of the electron density is efficient in these devices.

Cross-sections 6nm below the GaAs/AlGaAs interface of the self-consistent potential $V(y, z)$ for Hartree and LDA calculations are shown in Fig.~\ref{fig:LS_pot}.  When the wire is occupied, the potential is essentially flat, which shows that electron screening is efficient in this device.  The electron density along this same line is shown in Fig.~\ref{fig:LS_dens}.  As observed in the original paper, the exchange-correlation potential does not have a significant effect.  Laux \emph{et al.} saw oscillations in the electron density but these do not appear in our calculations.  Although we were not able to reproduce their modulations, discretizing the potential on a regular grid could have prevented screening at length scales similar to the lattice spacing.  We believe that the disappearance of the density modulations is due to the high accuracy potentials calculated using the FEA method.

\section{Electron Densities in the Low-Density, Weak-Confinement Regime}
\label{sec:phase_diagram}

We now turn our attention to the low-density, weak-confinement limit of the quasi-one-dimensional wire.  The split gate geometry from Sec.~\ref{sec:LS} does not allow the density and confinement of the wire to be tuned independently.  As the split gate voltage decreases, the transverse confining potential $V_{\mathrm{ext}} (\vec{r})$ at the GaAs/AlGaAs interface in the $y$-direction becomes weaker but its minimum decreases.  Electrons can flow into the system and the density increases.  The top-gate, split-gate device used in the series of experiments initiated by Hew \emph{et al.}~\cite{Hew:2009, Smith:2009, Kumar:2014} allows the confinement and density to be tuned separately.  The surface of the split gate device is covered with a layer of insulating polymer on top of which an additional metallic gate is deposited.  The voltage on the top gate $V_{tg}$ can be varied separately from the split gate voltage $V_{sg}$ such that the confining potential $V_{\mathrm{ext}} (\vec{r})$ can be weakened whilst keeping the electron density constant. In this section, we look at such a device which is described in more detail in App.~\ref{subapp:tgsg_device}.  The electron density was calculated using the effective-band approximation described in App.~\ref{app:effective_band} which produces very good approximate solutions to the Kohn-Sham equations.

Self-consistent solutions for this system were calculated for split gate voltages $V_{sg}$ in the range -0.58V to -0.46V and top gate voltages $V_{tg}$ in the range -1.75V to -1.10V.  This range of $V_{sg}$ corresponds to the weakest possible confining potential as the electrons are no longer depleted under the split gate for $V_{sg} > -0.46$V.  The range of $V_{tg}$ corresponds to the low density regime where the constriction is almost entirely drained of electrons.  It is in this regime that we expect electrostatic interactions and exchange-correlation effects to be enhanced.  In this section, we will describe the electronic structure of the quasi-one-dimensional quantum wire in this regime.

\begin{figure*}
  \subfigure[][\label{fig:Hartree_interconf} Hartree]{\includegraphics[width = 8.5cm]{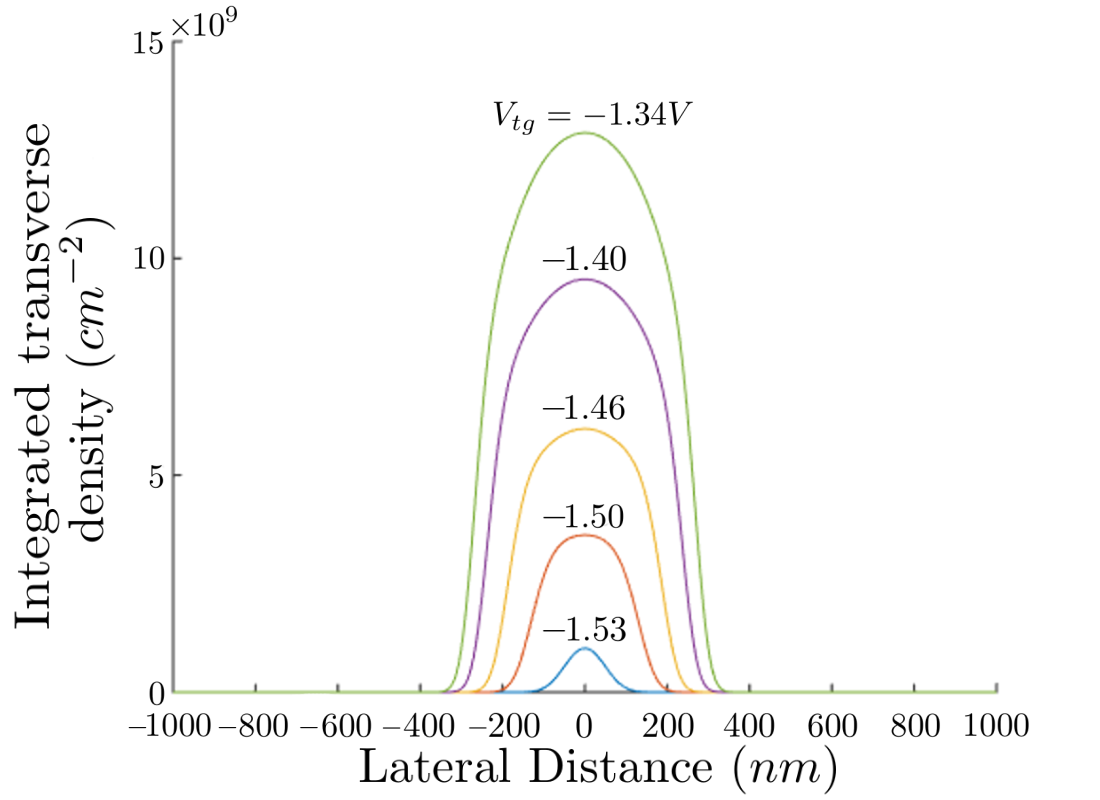}}
  \subfigure[][\label{fig:LDA_interconf} LDA]{\includegraphics[width = 8.5cm]{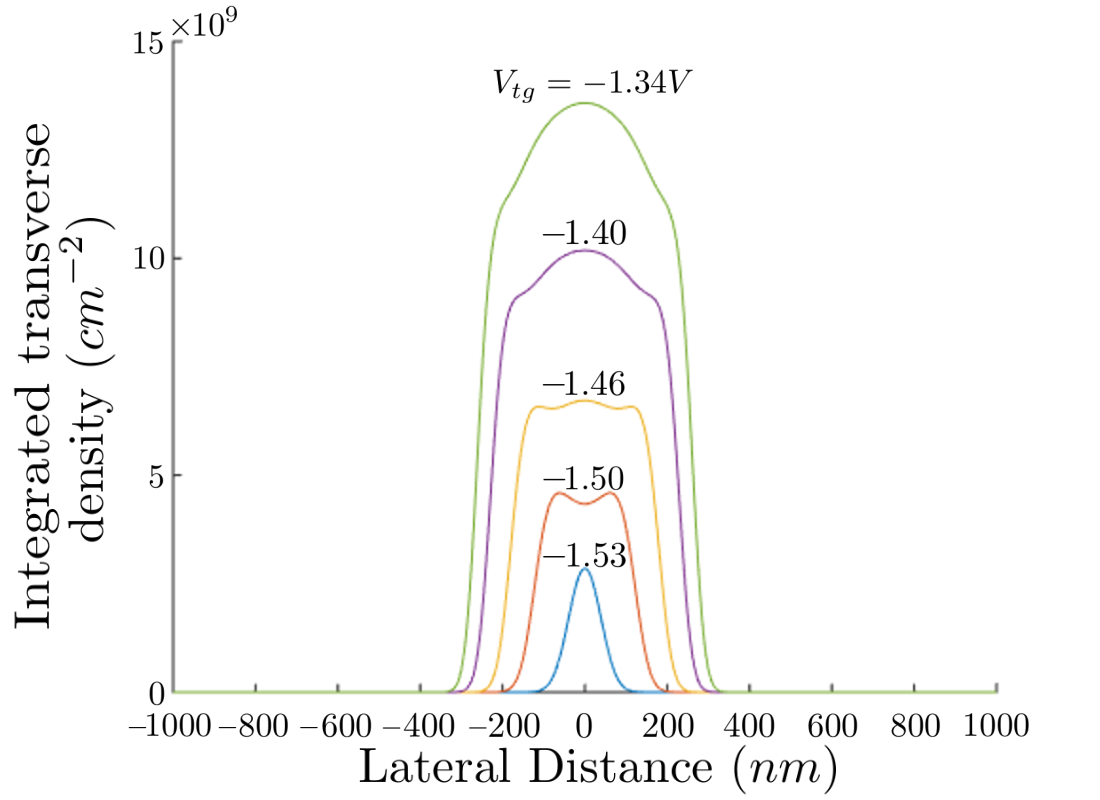}}
  \caption{(Color online) Integrated transverse density for $V_{sg} = -0.50$V at 50mK.  The electrons do not form a double-row density but the exchange-correlation potential generates small transverse density modulations which do not occur in the Hartree simulations.}
  \label{fig:interconf_dens}
\end{figure*}

For a fixed split gate voltage, as the top gate voltage becomes more negative and the density decreases, there is a qualitative difference induced by the exchange-correlation potential.  In the LDA calculations, we observed density modulations in the $y$-direction transverse to the wire axis which do not appear in the Hartree approximation, where only electrostatic interactions are taken into account.  A comparison of the different behaviours for Hartree and LDA calculations is shown in Fig.~\ref{fig:interconf_dens} for $V_{sg} = -0.50$V.  The varying top gate voltage changes the slope of the conductance band so the maximum of the electron density in the growth direction shifts.  Therefore, for the top-gate, split-gate devices simulated in this section, the density has been integrated over the $z$-direction.  For large densities, the electrons effective screen themselves but in the LDA calculations, as they are depleted from the wire, the modulations in the transverse density distribution appear.  The density modulations observed here are qualitatively different from those observed in the papers by Laux \emph{et al.}~\cite{Laux:1988} and Malet \emph{et al.}~\cite{Malet:2005} as they are produced exclusively by the exchange-correlation potential.

In the weak-confinement regime, for high densities the electrons also screen themselves effectively but as the electrons are depleted, the charge distribution localises to the edges of the wire and a double-row density distribution forms.  Due to the translational invariance of our simulations, this density distribution corresponds to both the zig-zag and double row phases of the weakly confined quantum wire~\cite{Mehta:2013}.  For most of the voltage values $V_{sg}$ and $V_{tg}$ where there are electrons in the wire, the external potential $V_{ext} (\vec{r})$ at the GaAs/AlGaAs interface has a single minimum in the transverse direction.  Row formation starts to occur in this regime where it is driven purely by the electrostatic repulsion of the electrons.  However, for the most negative values of $V_{tg}$, the top gate generates a double-well potential with two minima in the transverse direction.  This generates an additional anharmonic force on the electrons which is independent of the many-body effect.

\begin{figure*}
  \subfigure[][\label{fig:Hartree_weakconf} Hartree]{\includegraphics[width = 8.5cm]{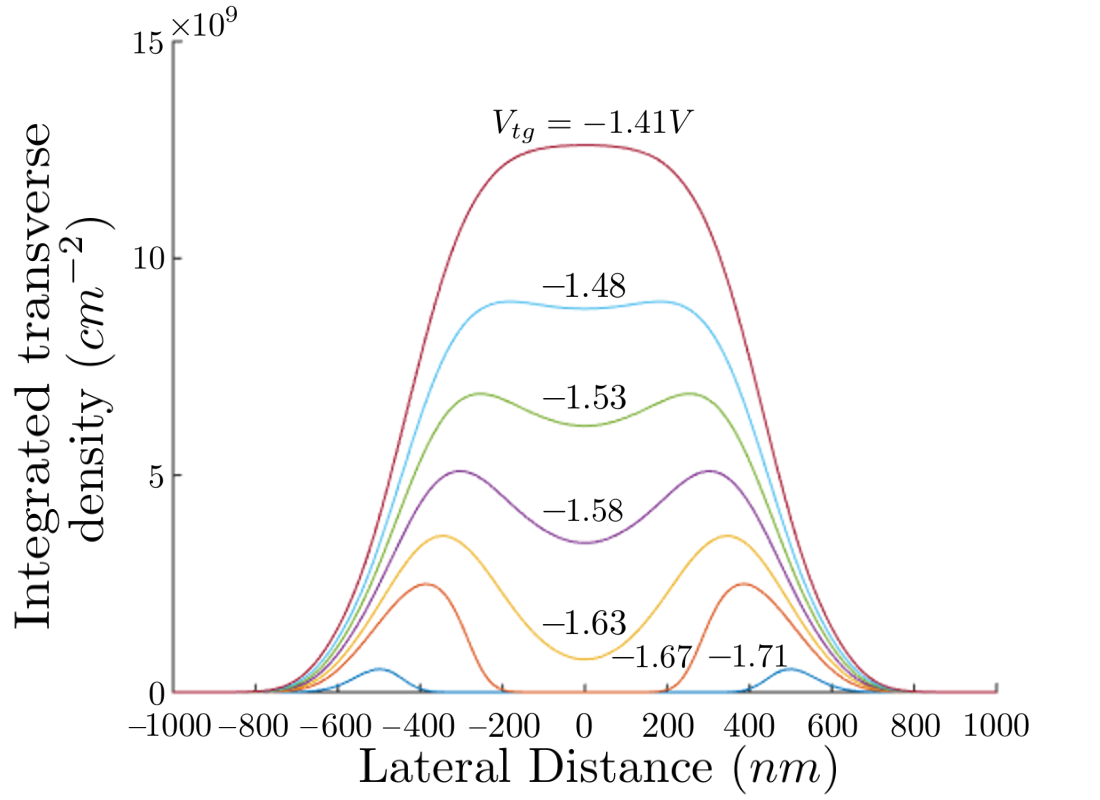}}
  \subfigure[][\label{fig:LDA_weakconf} LDA]{\includegraphics[width = 8.5cm]{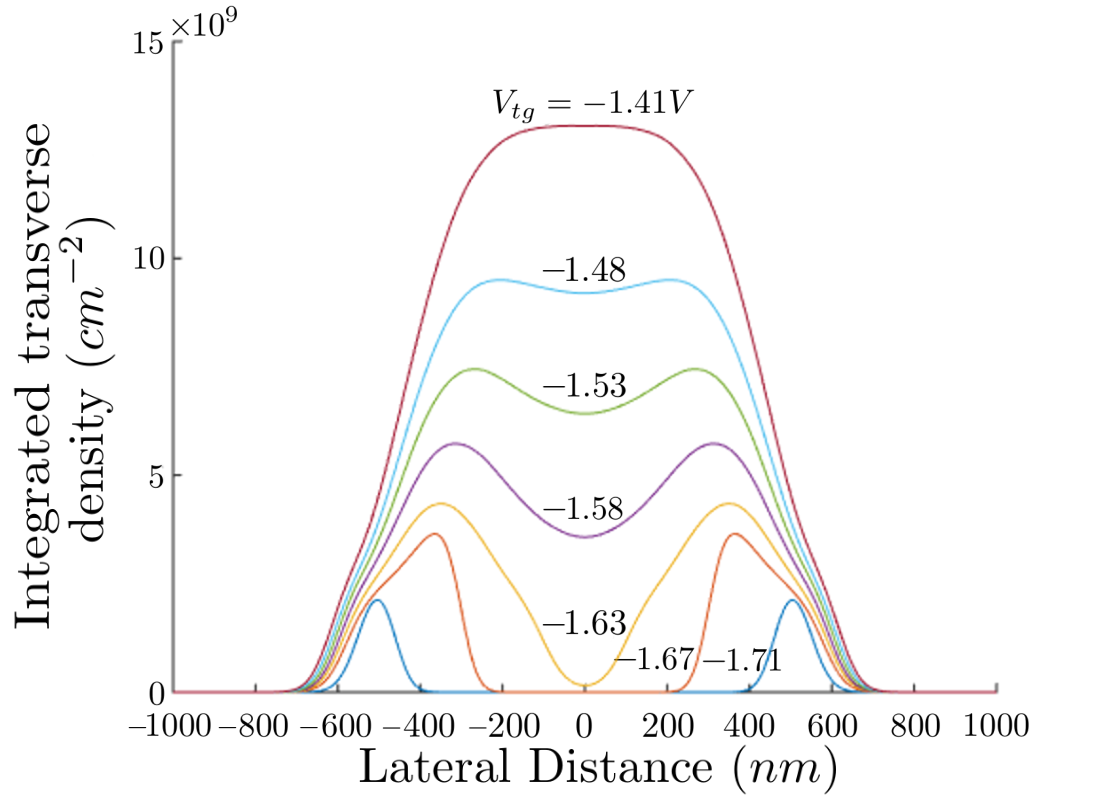}}
  \caption{(Color online) Integrated transverse density for $V_{sg} = -0.46$V at 50mK.  Traces show the density for different top-gate voltages from $V_{tg} = -1.70$V to $V_{tg} = -1.40$V in steps of $\Delta V_{tg} = 0.02$V.  This is the weak-confinement regime where electrostatic repulsion is dominant and double-row formation occurs.  For high densities, the electrons screen themselves but as the density decreases, two rows forms in both the Hartree and LDA calculations.  For $\Delta V_{tg} < -1.58$V, the external potential $V_{\mathrm{ext}} (\vec{r})$ acquires a double-well structure which enhances row separation.  When the wires are completely decoupled, the electron density lies almost entirely underneath the surface gates.}
  \label{fig:weakconf_dens}
\end{figure*}

A typical set of integrated transverse densities in this weak-confinement regime is shown in Fig.~\ref{fig:weakconf_dens} for $V_{sg} = -0.46$V.  Whilst there is some additional localisation for the LDA calculations, the main feature is the splitting of the electron density into two rows.  As the electron density is reduced, this is driven first by the electrostatic repulsion of the electrons and then by the external double-well potential.  Before the electrons are completely depleted from the wire, the two rows are practically decoupled and the device consists of a pair of quantum wires which lie beneath the surface gates.

\begin{figure*}
  \subfigure[][\label{fig:Hartree_nopks} Hartree]{\includegraphics[width = 8.5cm]{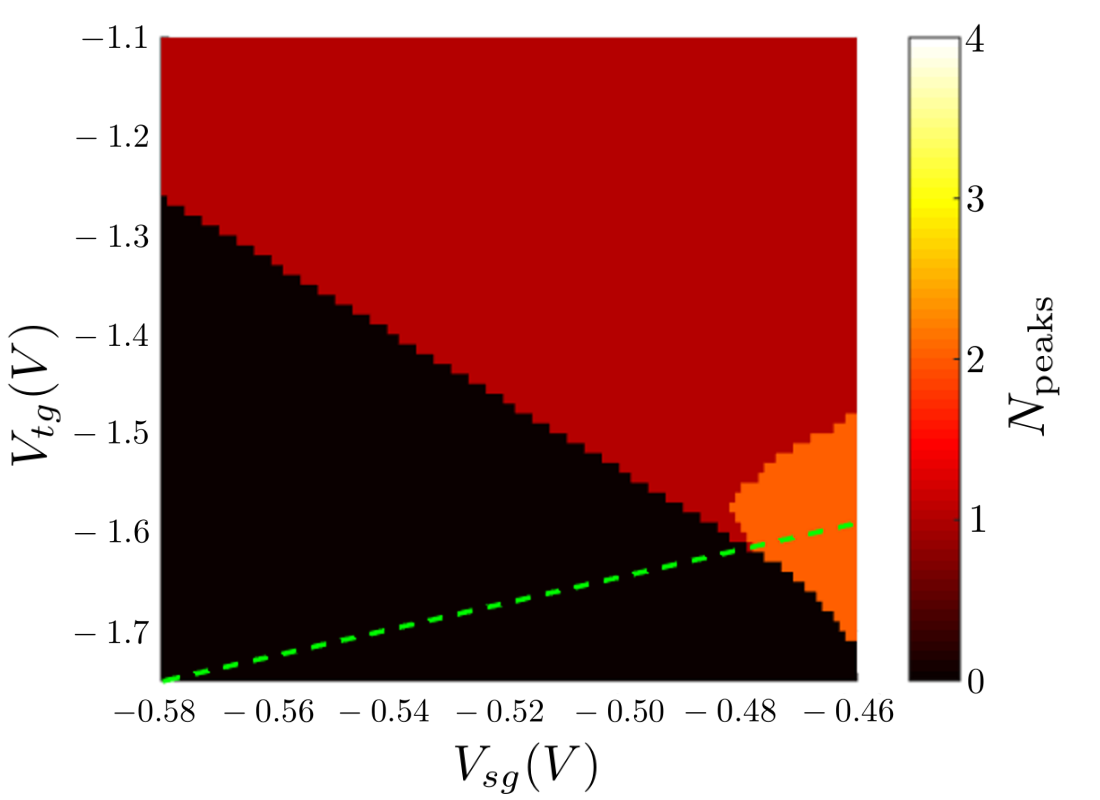}}
  \subfigure[][\label{fig:LDA_nopks} LDA]{\includegraphics[width = 8.5cm]{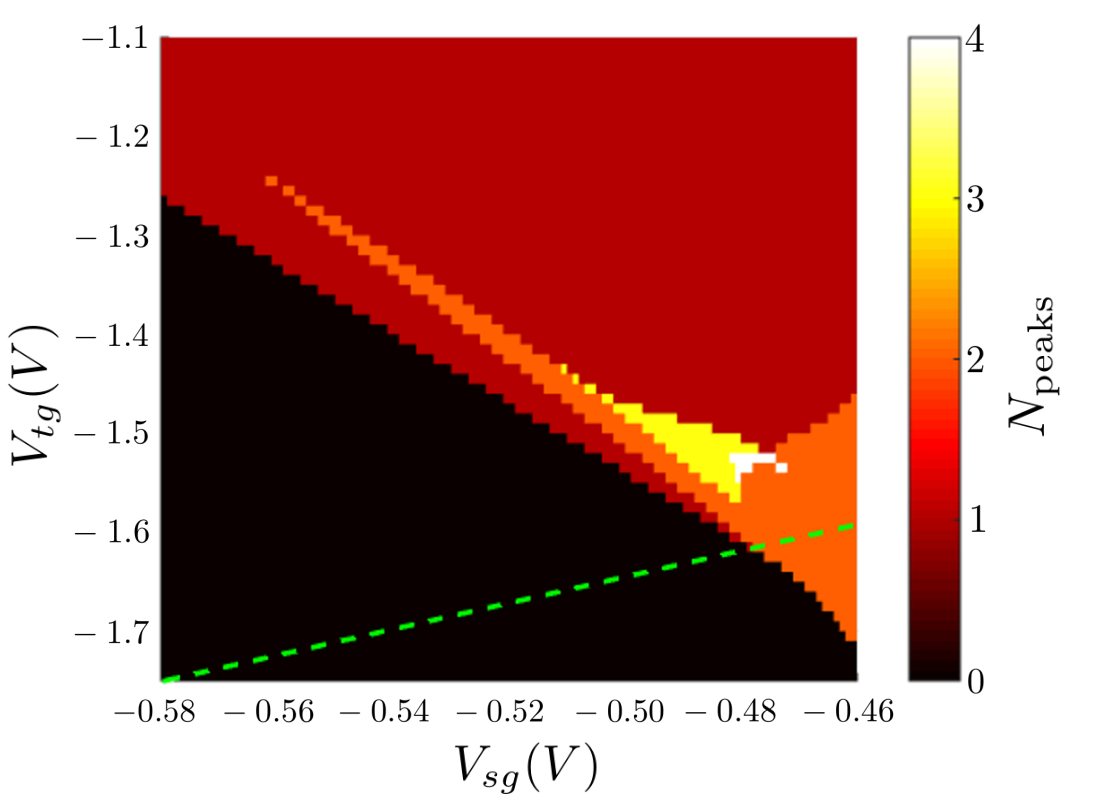}}
  \caption{(Color online) Number of peaks in the transverse density as a function of $V_{tg}$ and $V_{sg}$.  For weak confinement and low density, double-row formation occurs for small $V_{sg}$ and large $V_{tg}$ in both the Hartree and LDA calculations.  Transverse density modulations in the LDA calculations at low density create a peaked structure which extends to a less weakly confined regime where $V_{sg}$ is more negative.  Below the dashed green line, the $V_{\mathrm{ext}} (\vec{r})$ forms a double-well potential below the GaAs/AlGaAs interface.}
  \label{fig:no_peaks}
\end{figure*}

The number of peaks in the transverse density distribution is shown in Fig.~\ref{fig:no_peaks} as a function of $V_{tg}$ and $V_{sg}$.  The boundary where $V_{\mathrm{ext}} (\vec{r})$ just below the GaAs/AlGaAs interface becomes a double-well potential is denoted by the green dotted line.  We resolve the peaks in the density using a valley definition; two peaks exist when the minimum value of the density between the two peaks is less that a given factor of the maximum density, ie. $n_{min} < \beta n_{max}$.  This definition allows us to exclude spurious peaks when the density is approximately constant across the width of the wire.  For this paper, we have chosen $\beta = 0.99$.  The double-row regime is well defined, with a boundary which is similar for both Hartree and LDA calculations.  The transverse density modulations are observed for stronger confining potentials just before the quantum wire is depleted of electrons and only occur in the LDA calculations where the exchange-correlation potential is non-zero.  For large densities, the electrons screen themselves efficiently and there is no fine structure in the electron density distribution.

The voltages $V_{sg}$ and $V_{tg}$ determine the density of electrons and the confinement in the transverse direction but there is no simple relation between these voltages and the physical parameters of the quasi-one-dimensional electron system.  We can express the data in Fig.~\ref{fig:no_peaks} in terms of the two physical length scales of the system:  the Wigner-Seitz radius $r_s$, defined here as the average separation of the electrons in the longitudinal direction, and the distance in the transverse direction at which the electrostatic and confinement energies are the same $r_0$.  For a model harmonic potential of frequency $\omega$, $r_0 = (2 e^2 / \epsilon_0 \epsilon_r m^* \omega)^{\frac{1}{3}}$~\cite{Meyer:2009, Mehta:2013}. However, in our simulations, the electrostatic potential $V_{ext} (r)$ contains anharmonic contributions.  Therefore, we define $r_0$ as the distance for which
\begin{equation}
    \label{eq:r0_Definition}
    V_{ext}(r_0) = e^2 / 4 \pi \epsilon_r \epsilon_0 r_0
\end{equation}
For a double-well potential, $r_0$ is defined as the separation between the minima of the potential if this is greater than the value of $r_0$ calculated from Eq.~\ref{eq:r0_Definition}.  The lengths $r_s$ and $r_0$ are normalised by the effective Bohr radius for GaAs $a_B^*$.

\begin{figure}
    \includegraphics[width=\columnwidth]{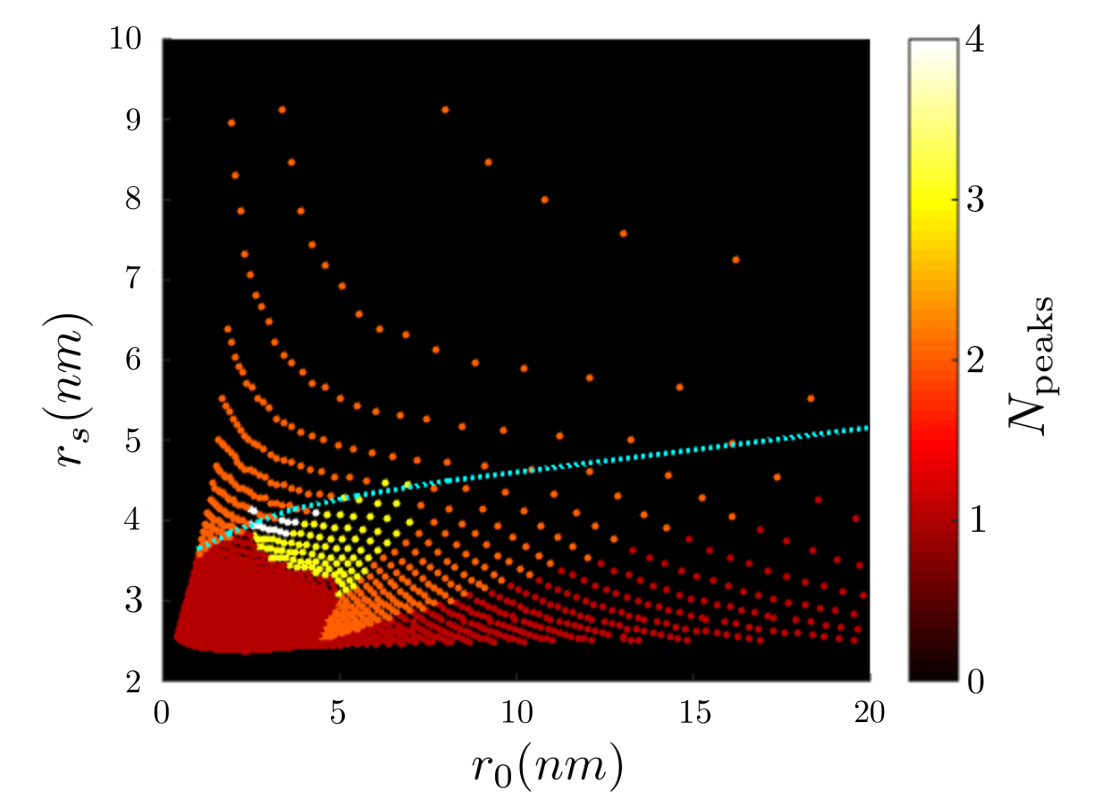}
    \caption{
    \label{fig:Deltar0rs}(Color online) Number of peaks in the transverse density as a function of $r_0$ and $r_s$ in the local density approximation.  A typical split gate devices, for example in Sec.~\ref{sec:LS}, corresponds to the small $r_s$ and $r_0$ limit.  The blue dotted line defines the transition above which the electrons separate into two rows in the Hartree approximation which corresponds to zig-zag or double row configurations and is due to electrostatic repulsion.  Modulations in the transverse density generated by the exchange-correlation potential occur for larger $r_s$ and $r_0$.  Calculations where $V_{\mathrm{ext}} (\vec{r})$ has two minima correspond to $r_0 > 10$ and are not shown in this figure.}
\end{figure}

The number of peaks in the transverse density distribution is plotted in Fig~\ref{fig:Deltar0rs} as a function of $r_s$ and $r_0$ for the LDA calculations.  We use the Hartree calculations to define the boundary between the exchange-correlation potential driven density modulations and double-row formation, which is denoted by the blue dotted line.  For small $r_0$, there is a strong confining potential and the density modulations are screened out in the Hartree approximation.  However, for the LDA, as the confining potential is weakened, it is possible to generate a multi-peaked density at low density.  For large $r_0$, the confining potential is weak and electrostatic repulsion forces the electrons into a double-peaked density distribution.  Such a density is consistent with a correlated system where the electrons are arranged in either a zig-zag configuration or in two parallel rows.  Interestingly, for lower densities the electrostatic transition to the double-row density occurs at weaker confining potentials, which is consistent with the quantum Monte Carlo calculations of Mehta \emph{et al.}~\cite{Mehta:2013}.

\begin{figure*}
  \subfigure[][\label{fig:Hartree_G} Hartree]{\includegraphics[width = 8.5cm]{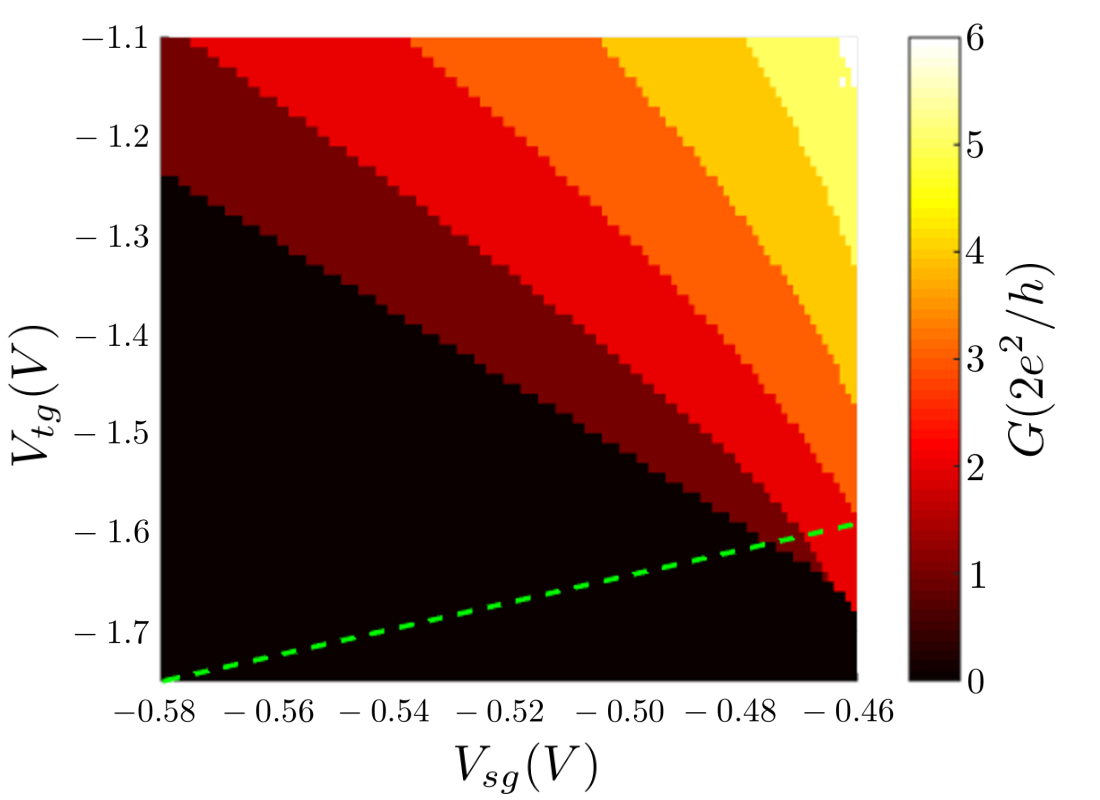}}
  \subfigure[][\label{fig:LDA_G} LDA]{\includegraphics[width = 8.5cm]{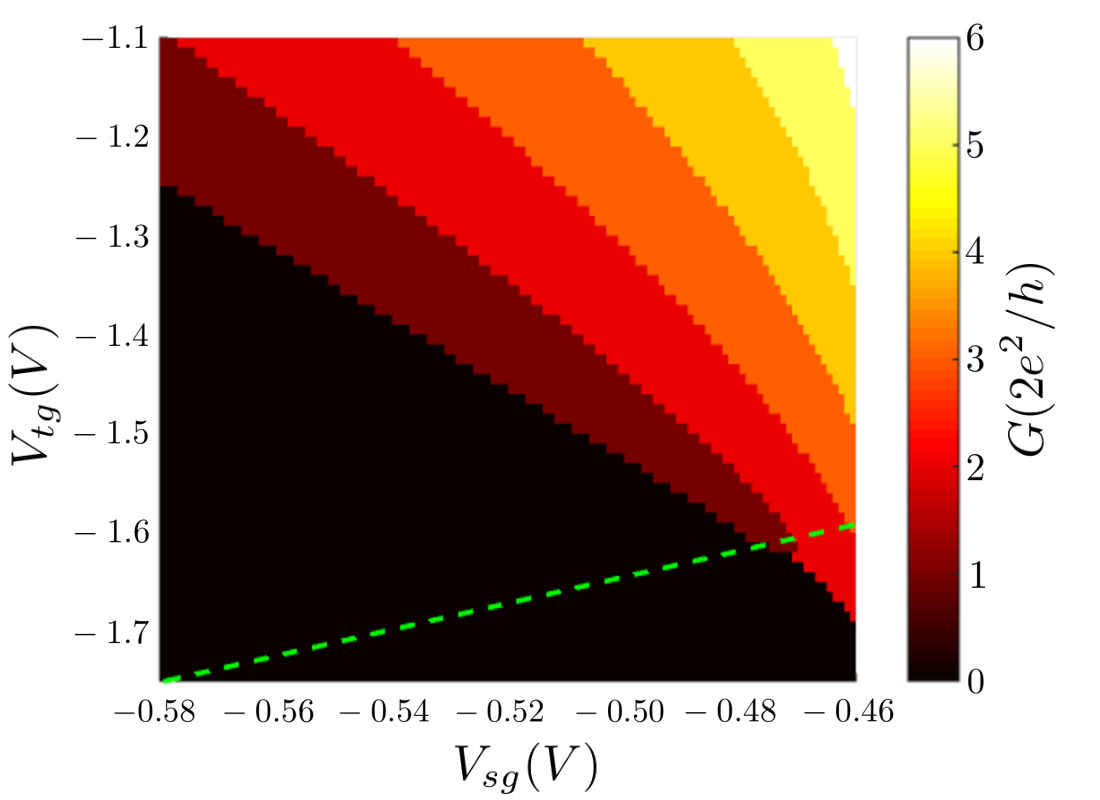}}
  \caption{(Color online) Number of Kohn-Sham bands below the chemical potential as a function of $V_{tg}$ and $V_{sg}$.  A double-jump in the number of conducting bands is seen for very weak-confinement and is driven by the formation of a double-well potential in the external gate potential $V_{\mathrm{ext}} (r)$ below the dashed green line.  The conduction bands do not correlate with either the formation of a double-row in the density or the density modulations generated by the exchange-correlation potential.}
  \label{fig:conductance}
\end{figure*}

The number of Kohn-Sham bands below the chemical potential is shown in Fig.~\ref{fig:conductance}.  In the Landauer-B\"uttiker formalism~\cite{DiVentra:2008, Buettiker:1990}, this is equivalent to the conductance of the quasi-one-dimensional wire $G$ in units of $2 e^2 / h$.  The plateaux edges are infinitely sharp as the wire is translationally invariant in the transport direction.  In both the Hartree and LDA approximations, we see that for a fixed top gate voltage $V_{tg} > -1.6$V, the conductance is quantised in units of $2 e^2 / h$ until the wire definition is lost at $V_{sg} = -0.46$V.  For more negative top gate voltages, we see a jump from zero to two occupied sub-bands.  As the electron rows separate, the coupling between the symmetric and antisymmetric Kohn-Sham bands tends to zero. This degeneracy leads to an initial jump in the conductance to $4 e^2 / h$ as the rows are decoupled and form two independent quantum wires.  In our simulations, there is no mechanism for breaking this degeneracy and regaining the missing plateau at $2 e^2 / h$ as seen in Refs.~\cite{Hew:2009, Smith:2009, Kumar:2014}.  A possible explanation for this effect would be a disorder-driven breaking of the parity symmetry of the system in the $y$-direction.  However, the inclusion of disorder, from the ionized dopant background for example, is outside the scope of this paper.

\section{Conclusions}
\label{sec:conclusions}

In this paper, we have studied the ground-state electronic structure of the quasi-one-dimensional quantum wire.  Self-consistent electron densities and electrostatic potentials for realistic semiconductor heterostructures were calculated with density functional theory using finite-element analysis.  We looked at the strong confinement regime by simulating a simple split-gate device, reproducing the simulations of Laux \emph{et al.}~\cite{Laux:1988}, using FEA to calculate accurate electrostatic potentials.  We saw that the transverse density modulations previously observed were screened out by the electron-electron interactions.

We extended these results into the weak-confinement, low-density regime by simulating a split-gate, top-gate device, which allowed us to control the density and transverse confinement of the quantum wire separately.  We showed that exchange-correlation effects produce small modulations in the transverse electron density for low electron densities across a range of confining potentials.  In the weak-confinement, low-density limit, the electrons separate into two rows due to electrostatic repulsion.  For the realistic experimental device we simulated, an external double-well potential forms which enhances this separation after row formation has occurred.  Decoupling of the rows leads to a degeneracy in the conducting states which produces an initial jump in the quantised conductance from zero to $4 e^2 / h$.  Our simulations provide no mechanism to explain the recovery of the first conductance plateau as the confinement of the quantum wire is further reduced~\cite{Hew:2009, Smith:2009, Kumar:2014}.

The modulations in the transverse density calculated in this paper are open to experimental investigation.  Using a weak magnetic field, electrons ejected from a quasi-one-dimensional wire can be focused into a quantum point contact (QPC).  Variations in the transverse density, such as the double rows observed in Fig.~\ref{fig:weakconf_dens}, would manifest themselves as variations in the current across the QPC as the magnetic field is varied~\cite{vanHouten:1989}.  Additionally, the tunneling of electrons from a quasi-one-dimensional wire into an adjacent two-dimensional electron system would contain signatures of the high-frequency transverse momentum components contained within the density modulations observed in this paper~\cite{Kardynal:1997, Macks:2000} as well as other correlation and many-particle effects~\cite{Jompol:2009, Grigera:2004, Altland:1999}.

Devices of this kind have the potential for many applications in quantum nano-electronics.   Electrostatic and Coulombic interactions allow  switching between one and two wires at length scales comparable to the surface gate features.  The ability to dynamically define quantum wires at the micron scale could be an important component in integrated quantum architectures.  For example, by overlaying the top gate over a small length of the split gate, a single wire could be split to form an Aharanov-Bohm ring with a tunable area, which could be used for high-precision magnetosensing.  The spontaneous localisation in the transverse direction demonstrated in this device could also be a precursor to Wigner crystallisation, which has applications in quantum information processing as a quantum memory or a spin-current mediator~\cite{Antonio:2015}.

Additionally, the one-dimensional wires created by the double-row formation lie almost entirely underneath the split gates.  In conventional one-dimensional quantum wires, electrostatic repulsion repels the electrons as far as possible from the split gate material.  However, sometimes it is desirable for the quantum channel to form closer to the surface gates.  For example, proposals to create Majorana fermions in quantum wires~\cite{Lutchyn:2010, Oreg:2010, Pientka:2013} typically rely on the proximity effect, where Cooper pairs can tunnel from a nearby superconductor into a one-dimensional wire which leads to bound electron pairs forming in the wire.  The strength of this effect is exponentially dependent on the distance over which the Cooper pairs need to tunnel.  The top-gate split-gate configuration studied here brings the electrons closer to the surface gate material and would enhance these effects, making it an ideal platform for studying proximity-effect induced physics in one-dimensional quantum wires.

\section{Acknowledgements}

The authors would would like to thank M. Pepper, J. T. Nicholls, C. J. B. Ford, I. Farrar, K. J. Thomas, S. Kumar, L. W. Smith and K.-F. Berggren for useful discussions.  We acknowledge the use of computing facilities from CamGrid.  This work was supported by the Engineering and Physical Sciences Research Council (EPSRC), UK, EP/K004077/1.

\appendix

\section{Calculating the Dopant and Surface Charges}
\label{app:dopant_calc}

The static charges consist of dopant and surface states which are thermally activated at high temperatures~\cite{Kumar:1993}.  At 70K, these states are frozen in their equilibrium configuration and the fraction of occupied states is fixed.  The resulting charge distribution provides the neutralizing background for the electrons in the quasi-one-dimensional quantum wire and must be calculated before performing the main DFT calculation.

We determine this static charge distribution using a one-dimensional self-consistent calculation at $T = 70$K.  The chemical potential of the GaAs surface is pinned in the middle of the band gap and we assume that it is also pinned at this value at the interface with the substrate.  When the gates are initially grounded, the system is translationally invariant.  For this simple one-dimensional calculation, Poisson's equation is solved on a regular grid with a lattice spacing of $\Delta z = 1$nm.  The energy of the dopant states is set to 5meV below the local conduction band minimum and the ionized dopant fraction is calculated using their thermal occupation.  The density of electrons in the conductance band is calculated in the Thomas-Fermi semi-classical approximation.  Convergence is achieved using the iterative method described in Sec.~\ref{sec:model} and the dopant density distribution at 70K is fixed.  The carrier density at the GaAs/AlGaAs interface is then calculated at the base temperature using DFT in the local-density approximation.  The number of active dopants can be calibrated so that the two-dimensional carrier density is the same as that which is experimentally measured for a specific semiconductor heterostructure.

As the semiconductor heterostructure is electrically neutral, the electric field above the surface of the sample vanishes.  Therefore, the charge per unit area on the surface must bend the potential such that $\vec{E} = - e \nabla \phi = 0$ for $z > 0$.  From Gauss' law, this requires that
\begin{equation}
    \sigma_{\mathrm{surface}} = e \epsilon_0 \epsilon_r (z = -0) \left. \frac{\partial \phi}{\partial z} \right|_{z = -0}
\end{equation}
at $T = 70$K where $\sigma_{\mathrm{surface}}$ is the charge per unit area of the surface.  Laux \emph{et al.}~\cite{Laux:1988} indicate that there should be a 20nm depletion region of the surface charges around the metallic surface gates.  We include this depletion region although its main effect on the quasi-one-dimensional quantum wire is to provide an offset in the gate voltages.

\section{Device structures}
\label{app:devices}

In this appendix, we detail the composition of the semiconductor heterostructures simulated in Sec.~\ref{sec:LS} and~\ref{sec:phase_diagram}.  For both devices, the wave functions are fully enclosed within the simulation domain and we found no evidence of edge effects.

\subsection{Split-gate device}
\label{subapp:LS_device}

The device simulated in Sec.~\ref{sec:LS} is the same as the one found in the simulations of Laux \emph{et al.}~\cite{Laux:1988}.  The device consists of two metallic surface gates at a voltage $V_g$ separated by a gap of 400nm on a semiconductor heterostructure with a 26nm GaAs cap above a 34nm $n$-doped layer of Al$_{0.26}$Ga$_{0.74}$As with a donor concentration of $N_D = 6.0 \times 10^{17}$cm$^{-3}$.  A 10nm Al$_{0.26}$Ga$_{0.74}$As spacer layer separates the dopants from a GaAs/AlGaAs interface against which the electrons are situated.  This GaAs layer extends to a substrate which is 6$\mu$m below the surface and contains a weak background $p$-doping with an acceptor density of $N_A = 10^{14}$cm$^{-3}$.  Above the sample is a 2$\mu$m layer of vacuum with the natural boundary condition $\partial_z \phi = 0$ on the top surface.  In these simulations, we set the density of charged surface states to $n_{\mathrm{surface}} = -1.6 \times 10^{12}$cm$^{-2}$ and we have included the 20nm depletion region of the surface charge adjacent to the gate metal, which gives a split gate voltage offset of +0.1V with respect to the original Laux~\emph{et al.} paper.  The electron density at the GaAs/AlGaAs interface was calculated for $T = 4.2$K on a regular grid with $\Delta y = 5$nm and $\Delta z = 1$nm from $y_{\mathrm{min}} = -200$nm to $y_{\mathrm{max}} = 200$nm from the center of the split gate and $z_{\mathrm{min}} = 90$nm to $z_{\mathrm{max}} = 64$nm below the surface.

\subsection{Top-gate split-gate device}
\label{subapp:tgsg_device}

The device modeled in Sec.~\ref{sec:phase_diagram} is the one used by Kumar \emph{et al.} in Ref.~\cite{Kumar:2014}.  The two metallic surface gates are separated by a gap of 800nm and the heterostructure consists of a 10nm GaAs cap above a 200nm Al$_{0.33}$Ga$_{0.67}$As $n$-doped layer with a 75nm Al$_{0.33}$Ga$_{0.67}$As spacer.  Below the spacer is the GaAs/AlGaAs interface against which the one-dimensional wire is defined.  The GaAs extends to a substrate 2$\mu$m below the surface and a 150nm layer of polymethylmethacrylate (PMMA) separates the surface from the top gate.  The donor concentration in the AlGaAs dopant layer was $N_D = 4.3 \times 10^{16}$cm$^{-3}$ which corresponds to a two-dimensional electron carrier concentration of $n_{2D} = 1.0 \times 10^{11}$cm$^{-2}$ at a base temperature of $T = 50$mK.  The electron density was calculated on a regular grid using the effective-band approximation described in App.~\ref{app:effective_band}.  The simulation regime for calculating the electron density was from $y_{\mathrm{min}} = -1000$nm to $y_{\mathrm{max}} = 1000$nm from the center of the split gate and $z_{\mathrm{min}} = -325$nm, $z_{\mathrm{max}} = -275$nm below the surface.  The lattice spacing of the grid was $\Delta y = 5$nm and $\Delta z = 1$nm.

\section{The Effective-Band Approximation}
\label{app:effective_band}

Solving the Kohn-Sham equation for a quasi-one-dimensional wire Eq.~\ref{eq:2D_Hamiltonian} is computationally expensive, especially in the weak-confinement regime studied in Sec.~\ref{sec:phase_diagram} where the wave functions can spread out in the $y$-direction.  To reduce the complexity of the problem, we used the effective-band approximation of Stopa~\cite{Stopa:1996}.  In this section, we show why this approximation works and the assumptions required to ensure its accuracy.

Let us expand the wave function in terms of the eigenstates of the Hamiltonian in the $z$ direction
\begin{equation}
    \label{eq:effective_band_wavefunction}
    \psi_j(y, z) = \sum_{n, y} a_{j, n} (y) \xi_n^y (z)
\end{equation}
where $\xi_n^y (z)$ are the solutions to the one-dimensional Schr\"{o}dinger equation
\begin{equation}
    \label{eq:effective_band_SE_z}
    \left( -\frac{\hbar^2}{2 m_*} \frac{\partial^2}{\partial z^2} + V(y, z) \right) \xi_n^y (z) = \epsilon_n (y) \xi_n^y (z)
\end{equation}
for fixed $y$.  As in Sec.~\ref{sec:model}, the density dependence of the exchange-correlation potential is dealt with iteratively so we will assume that $V(y, z)$ does not depend on $n(y, z)$ for the purposes of the approximation.  Inserting Eq.~\ref{eq:effective_band_wavefunction} into Eq.~\ref{eq:2D_Hamiltonian} for $k = 0$, premultiplying by $\sum_y {\xi_m^y}^* (z)$ and integrating over $z$ gives
\begin{equation}
    \label{eq:effective_band_SE_y}
   \left( -\frac{\hbar^2}{2 m_*} \frac{\partial^2}{\partial y^2} + \epsilon_m (y) \right) a_{j, m} (y) + f_{j, m}(y) = E_{j} a_{j, m} (y)
\end{equation}
where
\begin{align}
    f_{j, m}(y)  = & -\frac{\hbar^2}{2 m_*} \sum_{n, y} a_{j, n} (y) \intinfty {\xi_{m}^y}^* (z) \frac{\partial^2 \xi_n^y (z)}{\partial y^2} \dd z \nonumber \\
    \label{eq:HFResidual}
     & \quad -\frac{\hbar^2}{m_*} \sum_{n, y} \frac{\partial a_{j, n} (y)}{\partial y} \intinfty {\xi_{m}^y}^* (z) \frac{\partial \xi_n^y (z)}{\partial y} \dd z
\end{align}
Taking $y$ to be a parameter for the Hamiltonian in Eq.~\ref{eq:effective_band_SE_z}, Hellman-Feynman theory~\cite{Feynman:1939} allows us to rewrite the integrals in Eq.~\ref{eq:HFResidual} as
\begin{widetext}
\begin{eqnarray}
    \label{eq:Hellman_Feynman_I}
    \intinfty {\xi_{m}^y}^* (z) \frac{\partial \xi_n^y (z)}{\partial y} \dd z & = & \frac{\intinfty \partial_y V (y, z) {\xi_{m}^y}^* (z) \xi_n^y (z) \dd z}{\epsilon_{m} (y) - \epsilon_n (y)} \\
    \label{eq:Hellman_Feynman_II}
    \intinfty {\xi_{m}^y}^* (z) \frac{\partial^2 \xi_n^y (z)}{\partial y^2} \dd z & = & \frac{\intinfty {\xi_m^y}^* (z) \left( \partial_y^2 V(y, z) \xi_n^y (z) + 2 \partial_y V(y, z) \partial_y \xi_n^y (z) - 2 \partial_y \epsilon_n (y) \partial_y \xi_n^y (z) \right) \dd z}{\epsilon_{m} (y) - \epsilon_n (y)}
\end{eqnarray}
for $n \neq m$.  We want to show that the term $f_{j, m}(y)$ vanishes for the electrons in layered, gate-defined semiconductor heterostructures.  For Eq.~\ref{eq:Hellman_Feynman_I},
\begin{eqnarray}
    \intinfty \partial_y V (y, z) {\xi_m^y}^* (z) \xi_n^y (z) \dd z & \leq & \intinfty \modulus{\partial_y V (y, z) {\xi_m^y}^* (z) \xi_n^y (z)} \dd z \\
    & \leq & \sqrt{\intinfty \modulus{\partial_y V(y, z) \xi_n^y (z)}^2 \dd z} \\
    & = & \sqrt{\left< \modulus{\partial_y V(y, z)}^2 \right>_n}
\end{eqnarray}
where the second inequality is due to the Cauchy-Schwartz inequality and $\left< \cdot \right>_n$ is the expectation value of an operator for a state in the eigenfunction $\xi_n^y$.  We note that due to the orthogonality of the $\xi_n^y (z)$
\begin{equation}
    \modulus{\partial_y \xi_n^y (z)}^2 = \sum_k \modulus{\intinfty {\xi_k^y}^* (z) \frac{\partial \xi_n^y (z)}{\partial y} \dd z}^2
\end{equation}
so using the Cauchy-Schwartz inequality on the various terms in Eq.~\ref{eq:Hellman_Feynman_II} gives
\begin{equation}
    \intinfty {\xi_{m}^y}^* (z) \frac{\partial^2 \xi_n^y (z)}{\partial y^2} \dd z \leq \frac{\sqrt{\left< \modulus{\partial^2_y V(y, z)}^2 \right>_n}  + 2 \sqrt{\intinfty \modulus{\partial_y V(y, z) - \partial_y \epsilon_n (y)}^2 \sum_{k \neq n} \modulus{\frac{\sqrt{\left<\modulus{\partial_y V(y, z)}^2 \right>_n}}{\epsilon_n (y) - \epsilon_k (y)}}^2 \dd z}}{\epsilon_m (y) - \epsilon_n (y)}
\end{equation}
In semiconductor heterostructures, the electron wave functions are strongly confined in the growth direction so only the ground-state eigenfunction in the $z$-direction $\xi_0^y (z)$ is occupied and $a_{j, n} \approx 0$ for $n \neq 0$ and we only need $f_{j, 0} (y) \to 0$.  Any variations in the energy scales in the transverse direction along the heterostructure interface are orders of magnitude smaller than $\epsilon_1 - \epsilon_0$.  Additionally, for a typical heterostructure interface there are no degenerate states with $\modulus{\epsilon_n (y) - \epsilon_k (y)} \to 0$.  Therefore,
\begin{equation}
    \modulus{\epsilon_0 (y) - \epsilon_n (y)} \gg \sqrt{\left< \modulus{\partial_y V(y, z)}^2 \right>_n}, \ \sqrt{\left< \modulus{\partial^2_y V(y, z)}^2 \right>_n}, \ \partial_y \epsilon_n (y)
\end{equation}
\end{widetext}
so $f_{j, 0} (y) \to 0$ and Eq.~\ref{eq:effective_band_SE_y} reduces to a one-dimensional Schr\"{o}dinger equation with a band-dependent effective potential.  The argument made here for the accuracy of the effective-band approximation extends to the two-dimensional electron systems in the original paper by Stopa.  The density is given by
\begin{equation}
    n (y, z) = \sum_j \intinfty f (E, T) g_{1D}(E; \varepsilon_j) \modulus{a_{j, 0} (y)}^2 \modulus{\xi_0^y (z)}^2 \dd E
\end{equation}
This method significantly reduces the computational complexity of the electron density calculations which allows us to explore the low-density, weak-confinement regime.

\bibliographystyle{unsrt}
\bibliography{Row_Splitting_Phases}

\end{document}